# Effective Visualization and Analysis of Recommender Systems

Hao Wang

Haow85@live.com

Ratidar.com

**Abstract**—Recommender system exists everywhere in the business world. From Goodreads to TikTok, customers of internet products become more addicted to the products thanks to the technology. Industrial practitioners focus on increasing the technical accuracy of recommender systems while at same time balancing other factors such as diversity and serendipity. In spite of the length of the research and development history of recommender systems, there has been little discussion on how to take advantage of visualization techniques to facilitate the algorithmic design of the technology. In this paper, we use a series of data analysis and visualization techniques such as Takens Embedding, Determinantal Point Process and Social Network Analysis to help people develop effective recommender systems by predicting intermediate computational cost and output performance. Our work is pioneering in the field, as to our limited knowledge, there have been few publications (if any) on visualization of recommender systems.

**Index Terms**—recommender system, Takens Embedding, Determinantal Point Process, Social Network Analysis, visualization

## 1 Introduction

Recommender systems are ubiquitous in the internet industry. From late 1980's to 2022, there have been a tremendous amount of investment on the field. Lately, the focus of recommender system research has shifted from increasing accuracy metrics to a more comprehensive set of goals. Since 2017, AI fairness [1][2][3] has become the new buzz word in the field of recommender system research. This doesn't mean technical accuracy metrics are no longer the main concern. It's just a phenomenon as the consequence of years of efforts have already leading to satisfactory results on the accuracy metrics.

Recommender systems have evolved through different stages of development. The earliest recommender system algorithms are techniques such as collaborative filtering [4], matrix factorization [5][6], factorization machines [7], learning to rank [8][9], and hybrid shallow models [10]. 2016 [11][12][13] witnessed a surge in deep learning approaches in the top recommender system research venue – ACM RecSys. As a result, companies big and small all started to take on the new technological trend and produced effective recommender systems such as DeepFM [14], Deep Interest Network [15] and so on.

As more and more people joined the contest to achieve better technical accuracy, research topics such as fairness and diversity have attracted more and more attention. Another application field that is very important is context-aware recommendation [16][17]. As sensors turn more efficient and effective, there are more methods to collect contextual data for recommendation, which solves one of the biggest challenges of the topic – how to gather input data.

Improving performance of recommender system algorithms is the daily routine of many companies' AI departments. Companies like Baidu use a procedure called bad case analysis to analyze the reasons behind the bad performance and improves the product bit by bit. The need of effective tools for algorithmic analysis is urgent, but there has been very few publications on the topic.

Visualization powered algorithmic analysis has been a hot topic since the Researchers have agreed that effective visualization can greatly facilitate the development of big data products. However, data related to algorithmic analysis are heterogeneous. There does not exist a single elixir approach that can be used to explore the algorithmic structures of recommender systems.

For example, if we use Mean Absolute Error (MAE) as a metric, we need to find a method that can effectively analyze the 1-D data curve if Stochastic Gradient Descent approach is used and learning step is used as the x-axis variable. However, if we want to analyze the popularity bias effect, we need to visualize 2-D data. In the first case, we propose to use Takens Embedding [18] to elevate the data structures to higher dimensions. In the case of popularity bias effect analysis, we use heat map to visualize the 2-D data. There are more examples of using different techniques to solve different problems in this paper.

We find in our data analysis and visualization paradigm, we are not only capable of analyzing and visualizing technical accuracy metrics such as MAE, but also exploring other metrics such as fairness and diversity. Our approach is not only effective, but also comprehensive. To the best of our limited knowledge, we are among the first to propose a comprehensive set of data analysis and visualization tools to facilitate the design of recommender systems.

## 2 Related Work

Recommender system has different research subfields. One of the techniques that is versatile in quite many subfields is matrix factorization. The classic matrix factorization is designed to increase the technical accuracy metrics. Later, a more generic framework named SVDFeature [19] was invented to incorporate feature engineering in user-item feature decomposition, and thus greatly enhances the number of application contexts for the technique.

There have been different ways to optimize matrix factorization. One notable case is Alternating Least Squares [20] – a technique integrated in many different software packages such as Spark MLLib. Alternating Least Squares uses an alternating optimization approach to produce fast and reliable results.

In 2020, MatRec [21] was proposed as a special case of SVDFeature to solve the popularity bias problem based on matrix factorization framework. One year later, Zipf Matrix Factorization [22] and KL-Mat [23] were proposed to reduce the popularity bias effect using regularization terms specially designed for the problem.

Matrix Factorization variants are also suitable for cold-start problem. ZeroMat [24] was invented in 2021 as the first algorithm in history that solves the cold-start problem without extra input data. In 2022, another cold-start resolver named DotMat [25] was invented with superior performance and no extra input data. After experiments, the researchers discovered that not only these 2 algorithms could solve cold-start problems, but also alleviate the sparsity problem when used as a preprocessing step [26]. The hybrid models are named ZeroMat Hybrid and DotMat Hybrid, respectively.

Visualization of AI algorithms [27][28][29] has received a lot of attention in the field of InfoVis. Visualization of deep learning [30][31] models has been an extremely popular field over the years. However, emphasis on recommender system has not been enough. In this paper, we use the following techniques to analyze and visualize recommender system's algorithmic data.

Taken's embedding [18] was invented in 1981 to visualize low-dimensional data in higher dimension. We use the technique to visualize our data series in 1-D. Determinantal Point Process [32] inspired our analysis and visualization as well. Determinantal Point Process has been used to enhance diversity of recommender systems by Google Research and other institutions [33][34]. Heatmap [35] is also used by us and it has been applied elsewhere in InfoVis research to produce effective and beautiful visualization.

We also resort to social network analysis and visualization [36][37] in our publication. Social network analysis decomposes complex data structures into simpler structures that can better summarize the data. We apply techniques such as community detection [38] to segment the data set and visualize similarity matrix using different kinds of graph visualization algorithms [39].

### 3 Recommender System Benchmark

Matrix Factorization is one of the most successful recommender system paradigms for the past decade. The main idea behind the framework is to approximate the user item rating matrix with dot products of user feature vectors and item feature vectors. Precisely speaking, the loss function of the paradigm is as follows:

Notice that the framework reduces the space complexity of the user item rating values from O(mn) to O(k(m+n)) where m is the number of users, n is the number of items, and k is the dimension of user / item feature vectors.

Common optimization techniques used to solve the matrix factorization loss function for the optimal user/item feature vectors include Stochastic Gradient Descent (SGD), Adagrad, etc. Among the multitudes of different techniques, SGD is the simplest technique which takes a random sample of data points in computation of an incomplete form of gradients.

There have been a lot of variants of the matrix factorization approach. One notable contribution is SVDFeature, which models the user/item feature vectors as feature-based vectors. The framework is versatile and can be modified into different kinds of specializations that are widely applied in the industry.

One special example of matrix factorization is ZeroMat. ZeroMat is a milestone in the history of recommender systems. For the first time in the decades, ZeroMat solves the cold-start problem without extra input data. ZeroMat takes advantage of Zipf distribution and probabilistic matrix factorization, producing MAE results far superior to random placement, and only slightly inferior to the classic matrix factorization algorithm with historic user item rating values.

Another important invention is DotMat, which produces comparative recommendation results with classic matrix factorization. DotMat is also a cold-start resolution without extra input data. ZeroMat performs well on small random samples are selected during SGD. DotMat is more versatile as it suits large random sample size as well as small sample size.

ZeroMat and DotMat alleviate the sparsity problem as well as solving the cold-start issue. By using ZeroMat and DotMat as a preprocessing step to other algorithms, e.g., classic matrix factorization (ZeroMat Hybrid and DotMat Hybrid), researchers are able to achieve better MAE performance than single recommendation models.

One of the common heuristics used to tackle the cold start problem is random placement. Namely, we select random items for users when the user is new.

In our paper, we choose the following 7 recommender system algorithms for our data analysis and visualization: Classic Matrix Factorization, Random Placement, ZeroMat, DotMat, , DotMat Hybrid, user-based collaborative filtering, item-based collaborative filtering. Since ZeroMat and DotMat are comparatively new and lesser known in the community. We elaborate the algorithmic details in the following 2 sections.

### 4 ZeroMat

In this section, we focus on introducing a 2021 invention named ZeroMat. ZeroMat is the first cold-start algorithm in recommender system's history that solves the cold-start problem without using side information or extra data, in contrast with popular approaches such as meta learning.

ZeroMat assumes the user item rating follows the following distribution:

$$\frac{R_{i,j}}{R_{max}} \sim \frac{Rank_{max}}{Rank_{i,j}} \quad (1)$$

Then the algorithm plugs the distribution into the framework of probabilistic matrix factorization:

$$P(U, V \mid R, \sigma_U, \sigma_V) = \prod_{i=1}^{N} \prod_{j=1}^{M} (U_i^T \cdot V_j) \times \prod_{i=1}^{N} e^{-\frac{U_i^T \cdot U_i}{2\sigma_U^2}} \times \prod_{j=1}^{M} e^{-\frac{V_j^T \cdot V_j}{2\sigma_V^2}} \quad (2)$$

The SGD (stochastic gradient descent) update formulas for the algorithm is as follows (with standard deviations simplified to a constant):

$$U_i = U_i + \gamma \times \left(\frac{V_j}{U_i^T \cdot V_j} - 2 \times U_i\right) (3)$$

$$V_j = V_j + \gamma \times \left(\frac{U_i}{V_j^T \cdot U_i} - 2 \times V_j\right) (4)$$

As can be observed from the update formulas, there is no extra information involved in the computation other than the parameters U and V.

## 5 DOTMAT

DotMat is the second invention in recommender system's history that solves the cold-start problem without input data. The algorithm was invented in 2022 – one year later than ZeroMat. It achieves even better results than ZeroMat.

The algorithm of DotMat was inspired by ZeroMat and RankMat. It modifies the loss function of the classic matrix factorization in the following way:

$$L = \left|(U_i^T \cdot V_j)^{U_i^T \cdot V_j} - \frac{R_{ij}}{R_{max}}\right| (5)$$

Just like ZeroMat, DotMat's SGD update formulas do not rely on extra information in its update formulas, and hence is history independent.

## 6 TAKENS EMBEDDING

In this paper, we propose to use Takens Embedding to analyze and visualize the MAE curve. MAE curve is the men absolute error curve for recommender systems, namely the absolute value of the error between the prediction and ground truth. Takens Embedding was proposed to elevate low dimensional datasets to higher dimensions while preserving the chaotic properties of the original data. Takens Embedding is one of the tools in the toolkit of topological data analysis.

The formal definition of Takens Embedding can be found in the original 1981 paper:

Let $M \subset \mathbb{R}^n$ be a compact manifold of dimension n. Let

$$\varphi: R \times M \to M$$

and

$$f: M \to R$$

be generic smooth functions. Then, for any τ>0, the map

$$M \to R^{2n} + 1$$

defined by

$$x \mapsto (f(x), f(x_1), f(x_2), \ldots, f(x_{2n}))$$

Where

$$x_i = \varphi(i \cdot \tau, x)$$

is an injective map with full rank.

In our research, we loosen the requirement on map M to allow embedding existing in dimensions lower than 2n. To our surprise, this slight modification of Takens Embedding leads to effective analysis and visualization of data series in 2-D space. We also loosen the requirement that M should be smooth. Since most data series in algorithmic results are continuous rather than smooth, we only require M to be continuous.

Takens embedding was originally invented to reconstruct attractors of dynamical systems in high dimensions. As shown in the next section, the algorithmic result curve of algorithms does not contain attractors, but the elevation of data from 1-D to 2-D and 3-D makes it possible to visualize and analyze the data more clearly.

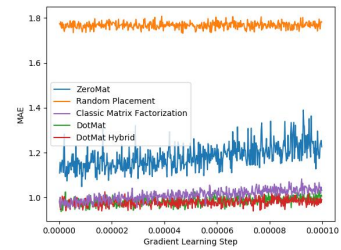

Fig. 1 MAE curves of 5 different recommender systems on MovieLens 1 Million Dataset

Fig.1 shows the MAE curve of 5 different recommender systems on MovieLens 1 Million Dataset [40]. We use Takens Embedding to visualize the curves in higher dimensions :

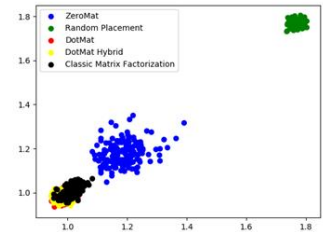

Fig.2 2-D visualization of dataset in Fig.1

Fig.2 illustrates the result of visualizing MAE curves in 2-D space. It is apparent that ZeroMat has the largest data point span, while random placement and DotMat are most compact. DotMat, DotMat Hybrid, and classic matrix factorization are clustered together, which means they probably share similar properties. We can safely draw the conclusion that these 3 algorithms produce the best performance when it comes to robustness.

After elevating the dimension of MAE curves into 3-D, by careful observation we draw the same conclusion as in 2-D. The visualization in 2-D and 3-D is clearer and more effective than the cluttered lines in 1-D space. The span and skewness of the MAE values are more visible.

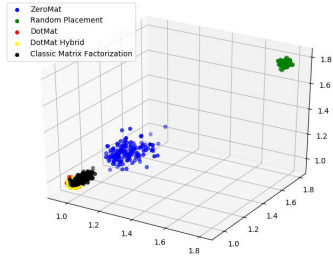

Fig.3 3-D visualization of dataset in Fig.1

We now visualize MAE curves of LDOS-CoMoDa dataset. LDOS-CoMoDa dataset [41] is a movie dataset including contextual information. LDOD-CoMoDa include 121 users and 1232 movies.

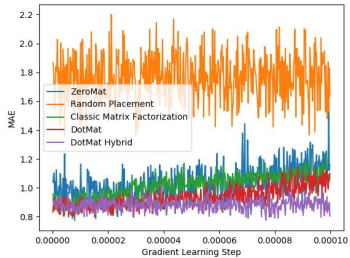

Fig.4 Visualization of MAE curves on LDOS-CoMoDa dataset

Fig. 4 demonstrates the MAE curves of 5 different recommender systems. It is very difficult to analyze 4 of them since they are cluttered together.

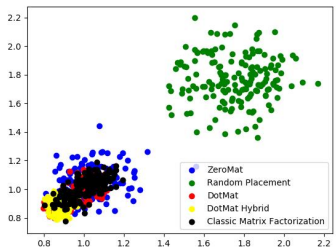

Fig. 5 illustrates the 2-D visualization of the MAE curves.

We elevate the dataset into 2-D visualization in Fig. 5.

Although 4 algorithms are still cluttered, but they are more visible and analyzable in point cloud format. DotMat Hybrid has the smallest diameter, with DotMat coming second in diameter length. Classic matrix factorization is also very compact. ZeroMat is much less compact and random placement is the most spread-out.

We now elevate 1-D MAE curves into 3-D space :

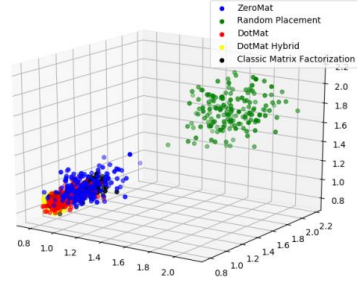

Fig. 6 illustrates the 3-D visualization of the MAE curves

In 3-D space we could explore the point clouds interactively, so we could analyze the data even more easily. Unlike in 1-D time series, we could examine the data in different aspects. In addition, unlike the 1-D MAE curve, in 3-D we can examine individual data point more clearly and easily as we can rotate and zoom the data set. This makes it a lot easier to detect abnormal data points or special structures.

## 7 RECURRENCE PLOT

Recurrence plot is a technology invented to visualize dynamical system recurrence patterns. In this paper, we use recurrence plot to show the recurring structures of the MAE curve.

Recurrence plot is a 2-D image defined as follows :

$$I(x,y) = \begin{cases} 1, & if\ |T(x) - T(y)| < \varepsilon \\ 0, & if\ |T(x) - T(y)| > \varepsilon \end{cases}$$

, where T denotes the time series (MAE curve, in our case), and $\varepsilon$ is a small real number.

We use recurrence plot to visualize MAE values of 5 different recommender systems with grid search on different gradient learning steps:

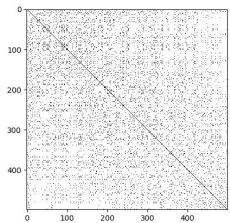

Fig. 7 Recurrence Plot of ZeroMat

From Fig. 7 to Fig. 10, we observe that DotMat Hybrid is the most robust algorithm since the black dots representing 1 are so densely populated in the graph. Random Placement is by theory and observation produces the most random result. DotMat as a single model has correlated structures, but it also looks like there is some redundancy in the graph since some areas are densely populated by black dots while others are nearly entirely blank.

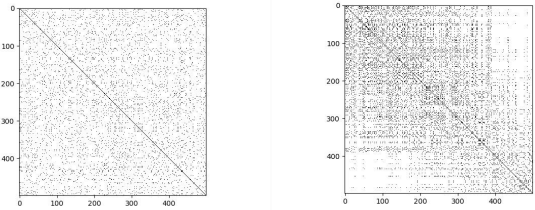

Fig. 8 & Fig. 9 Recurrence Plot of Random Placement and DotMat

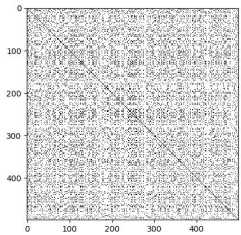

Fig. 10 Recurrence Plot of DotMat Hybrid

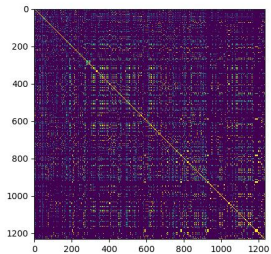

Fig. 11 HeatMap of user-user similarity

## 8 DETERMINANTAL POINT PROCESS

In modern day recommender systems, diversity is one of the critical concerns of product owners. One way to enhance diversity is to penalize the loss function of recommender systems using a regularization term computed by Determinantal Point Process (DPP).

DPP constructs the similarity matrix of items, and then use the determinant of the matrix as a regularization term to the loss function of recommender system algorithms. The maximization of the determinant is equivalent to maximizing the volume spanned by vectors of the similarity matrix. The idealized maximum value of the volume is achieved when the spanning vectors are orthogonal to each other.

To analyze and evaluate the diversity of recommender systems, we use heatmap to plot the similarity scores between item pairs before and after the execution of our algorithms. We define similarity as in collaborative filtering algorithms. Fig. 11 and Fig. 12 demonstrate the similarity heatmaps computed on the LDOS-CoMoDa Dataset. Fig.11 illustrates item-item similarity, showing that the items are less effected by popularity bias compared with user-user similarity result shown in Fig. 12. This suggests that we should apply item-based collaborative filtering rather than user-based collaborative filtering. Also we should do more item-based similarity computation than user-based similarity computation.

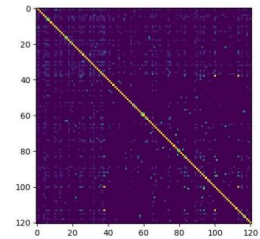

Fig. 12 HeatMap of item-item similarity

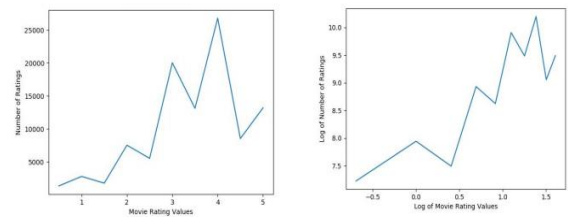

Fig.13 Visualization of Popularity Bias Effect on MovieLens Small Dataset

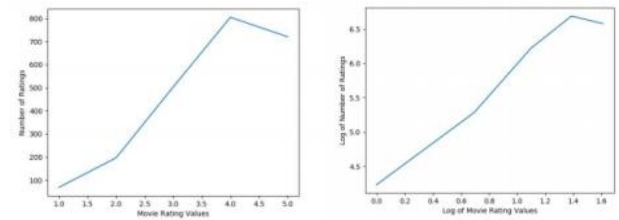

Fig.14 Visualization of Popularity Bias Effect on LDOS-CoMoDa Dataset

## 9 POPULARITY BIAS AND LONG-TAIL EFFECT

Bias problem is one of the hottest research topics of recommender systems in recent years. Both industry and governmental agencies carried out a series of experiments to enhance the transparency and fairness of recommender systems including the efforts to alleviate popularity bias problems.

Popularity bias refers to the phenomenon that the most popular items of a recommender system effects the performance far more deeply than others. For example, in the collaborative filtering algorithms, the most popular items are involved in far more similarity computations than the rest.

In 2018, Wang et. al. [42] designed analytical formulas to capture the effect of popularity bias in the input structures on algorithmic performance. In their paper, the researchers prove that Zipf Law in the input data structures leads to power law effect in intermediate computational procedures. In this paper, the authors plot graphs of item ranks v.s the number of items to illustrate the popularity bias.

Borrowing the ideas from their work, we plot the graph of item user item rating values v.s the number of ratings in the input data structures of MovieLens 1 Small Dataset in Fig. 13 (Log-log plot). From the figures, we observe that the user item ratings follow stepwise power law distribution. Therefore it is safe to apply algorithms such as ZeroMat who makes the same assumption for the dataset. Visualization of LDOS-CoMoDa dataset in Fig.14 leads to analogous analysis. Our visualization helps us to choose algorithms to tackle the problem.

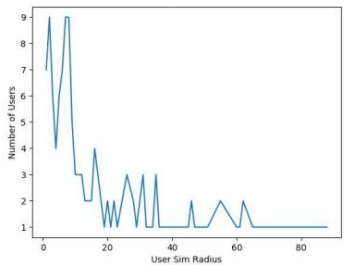

Fig.15 Visualization of user similarity radius on LDOS-CoMoDa Dataset

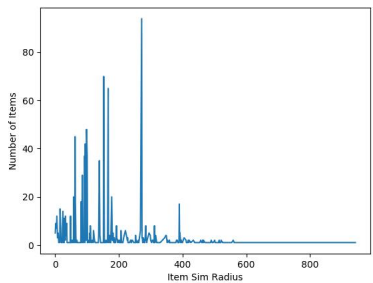

Fig.16 Visualization of item similarity radius on LDOS-CoMoDa Dataset

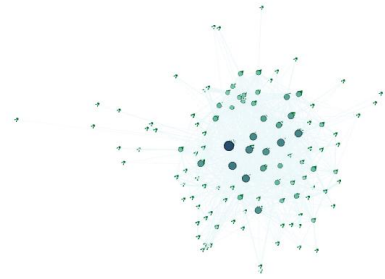

Fig.17 User similarity matrix visualization on LDOS-CoMoDa dataset

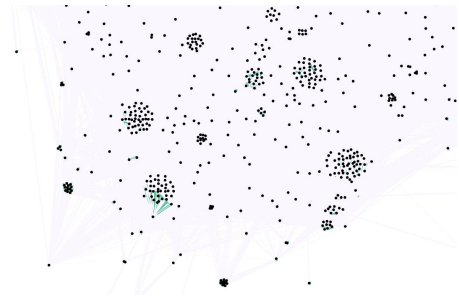

Fig.18 Local view of item similarity radius on LDOS-CoMoDa dataset

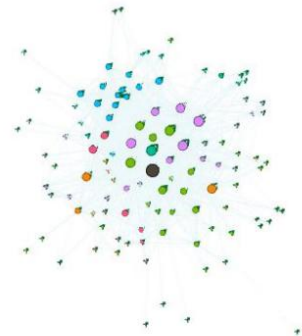

Fig.19 Community detection of user similarity radius on LDOS-CoMoDa dataset

## 10 Visualization of Similarity Radius

We propose a concept named Similarity Radius in this paper. A similarity radius of a user is defined as follows:

Definition 9.1: Similarity Radius of a user is the number of users whose similarity score with her is greater than 0.

Analogously, similarity radius of an item is defined as follows:

Definition 9.2: Similarity Radius of a user is the number of users whose similarity score with her is greater than 0.

Similarity Radius computes the size of the neighborhoods of users and items. The larger the similarity radius is, the more impact the user/item will exert on the intermediate computational procedure. We plot the user/item popularity ranks v.s similarity radius to explore the relations between popularity bias and intermediate similarity computational costs.

Fig. 15 shows the user-user similarity radius of LDOS-CoMoDa Dataset, and Fig. 16 illustrates the item-item similarity radius of LDOS-CoMoDa Dataset.

Similarity radius measures the popularity bias effect in the intermediate computational procedures of recommender systems such as collaborative filtering. Collaborating filtering, among many algorithms, needs to compute the similarity between users or items. The similarity radius determines the number of computational steps in the intermediate procedure. If similarity radius is unevenly distributed, MapReduce will suffer skewness problem. In addition, this will effect the performance of algorithms as well.

From Fig.15, we observe that user similarity radius is highly skewed and even exhibiting power-law effect. This probably means we should prefer item-based similarity computations (Fig.16) which is much more evenly distributed. The visualization helps us detect potential intermediate computational problems in computational procedures of recommender systems.

## 11 Visualization by Social Network Analysis

The similarity matrix discussed in previous sections can be visualized in different ways other than heatmaps. If we take users / items as data points in social network graphs, and user-user / item-item similarity pairs as edges, we acquire a social network built-up the similarity matrix and we could all sorts of social network analysis (SNA) techniques to investigate the data.

We apply social network visualization to similarity pairs generated on LDOS-CoMoDa datasets and obtain Fig. 17 and Fig. 18. Fig.17 illustrates the user-user similarity graph, the size of whose nodes represents the similarity radius. Fig.18 illustrates a local view of the item-item similarity graph. As can be observed from Fig. 18, there exist many small clusters in the graph.

Comparison between Fig. 17 and Fig. 18 leads us to believe that item-based collaborative filtering might be more effective in computational time than user-based collaborative filtering since we can compute similarities within small clusters to expedite our overall computational process.

We could also apply other social network analysis and visualization algorithms to the similarity matrix. Fig. 19 shows the community detection result of LDOS-CoMoDa dataset using Louvain's Method.

## 12 Conclusion

In this paper we provide systematic visualization and analysis of 7 different recommender algorithms and 2 open-source datasets. We demonstrate that by carefully selecting visualization techniques, we are able to predict intermediate computational cost and output performance, therefore choosing correct recommender systems beforehand.

In future work, we would like to explore visualization of other AI algorithms to help algorithm engineers and experts design new algorithms and improve old ones. We believe AI + visualization can transform the current IT industry and research community into a more advanced technological landscape.